\documentclass[%
 reprint,
 amsmath,amssymb,
 aps,
prl,
]{revtex4-2}

\usepackage{graphicx}
\usepackage{dcolumn}
\usepackage{bm}
\usepackage[dvipsnames]{xcolor}
\usepackage{hyperref}
\usepackage{orcidlink}

\begin{document}

\preprint{APS/123-QED}

\title{Hollow Beam Optical Ponderomotive Trap for Ultracold Neutral Plasma}

\author{S. A. Saakyan\,\orcidlink{0000-0001-8870-5336}}
\affiliation{Joint Institute for High Temperatures, Russian Academy of
	Sciences, Moscow 125412, Russia}
\email[]{saakyan@ihed.ras.ru}

\date{\today}

\pacs{32.80.Ee, 32.60.+i, 34.20.Cf, 51.30.+i, 64.60.De}

\begin{abstract}
	Rapidly oscillating, inhomogeneous electromagnetic field from laser exert a 
	force that repels charged particles from regions of high light intensity. 
	We propose and analyze a flat-bottomed hollow-beam ponderomotive optical trap 
	for an ultracold neutral plasma (UNP), driven by a high-power CO$_2$ laser.
	Molecular dynamics simulations show that the plasma and Rydberg atoms are 
	effectively trapped within a nearly uniform dark region bounded by repulsive 
	light walls.
	In contrast to RF traps, flat-bottomed traps yield a small density-weighted mean
	ponderomotive energy per electron, while the UNP collision frequency is far 
	below the laser frequency, thereby making collisional absorption negligible and 
	does not limit the lifetime of the trap.
	This approach could enhance antimatter production and storage.
\end{abstract}

\maketitle
Ponderomotive confinement is the opposite effect of the well-known ponderomotive
acceleration of charged particles~\cite{PonderomotiveAcceleration2001,he2003ponderomotive,
	PlasmaAcceleratorsRev2009}. 
The ponderomotive force is proportional to the gradient of the cycle-averaged field intensity and repels charged particles from regions of high field amplitude~\cite{JETP1958, ElTrap1966}.
Ponderomotive traps (PTs) are widely used to confine non-neutral, single-component 
plasmas, such as ions or dusty particles in RF traps~\cite{Dubin1999, fortov2004dusty, IonTrap2003, ElectronTrap2021}, as well as electrons in ponderomotive optical potentials~\cite{ElTrap1966, moore1992confinement, chaloupka1997single, chaloupka1999observation}, and Rydberg atoms in bottle-beam traps~\cite{PRLPTrapforRy3D2020,TrapRydberg2023} or ponderomotive lattices~\cite{ponderLattRy2000, anderson2011trapping}.
By contrast, confinement of quasi-neutral plasmas in PTs is limited by collisional absorption (inverse bremsstrahlung), which rapidly heats the electrons in an external oscillating field~\cite{silin1965, mulser2010high, LuitenHeating2012}.
These limitations can be bypassed using optical box or hollow-beam traps. 
Recent advances in light shaping have enabled the creation of sculpted optical 
potentials for neutral-atom confinement~\cite{NarutePhReviw2021}. 
Blue-detuned optical boxes use optical-dipole forces to confine atoms in dark, flat-bottom regions, suppressing photon scattering and ac-Stark shifts and yielding uniform-density samples.
Applying a flat-bottom hollow-beam trap to an ultracold neutral plasma (UNP)~\cite{killian2007ultracold,LaserCoolingUCP2019,gorman2021magnetic,PRLSteadyState2024, bergeson2022ultracold} minimizes the density-weighted mean ponderomotive energy per electron by analogy with neutral-atom dark boxes, where scattering is suppressed.
Additionally, because the optical drive frequency is much larger than the electron–ion collision rate in UNPs, collisional absorption is negligible on trapping timescales~\cite{IBfreqcalc2024,IBfreqexp2023}.

In this Letter, we propose and analyze a novel UNP confinement method using 
the dark core of a hollow laser beam via molecular dynamics (MD) simulations.
Simulations show that a two-component lithium plasma and Rydberg atoms are confined 
within a nearly uniform dark core bounded by repulsive light walls. 
We identify optimal experimental parameters and show
that the trap simultaneously stores high-density UNP clusters and
Rydberg states formed by three-body recombination, enabling dual
trapping of UNP and Rydberg atoms.
These capabilities suggest potential applications for trapping 
antiproton-positron plasmas and producing high-density positronium 
samples~\cite{antihydrogenSpectroscopy, positronPRL, steinbrunner2023thermal}.

The ponderomotive force $\mathbf{F}{_\text{p}}=-\mathbf{\nabla}U{_\text{p}}$
acts on a free charged particle in a rapidly oscillating, inhomogeneous 
electromagnetic field, where 
$U_\text{p}=e^2(2m_{\text{e,i}}\omega^2c\varepsilon_0)^{-1}I(r)$
is the ponderomotive energy of the particle~\cite{zhang2011magic}. 
Here, $e$ is the elementary charge, $m_{\text{e,i}}$ is the mass of the 
electron or ion, $\omega$ is the angular frequency of the optical field, and 
$I(r)$ is the optical field intensity distribution.

LG$_{0\ell}$ is an example of a laser beam with a minimum on the 
axis~\cite{kuga1997novel}. 
The intensity distribution of the hollow beam at the waist is given by
\begin{equation}\label{e_LGbeam}
	I(r)=P_0\frac{2^{\ell+1}r^{2\ell}}{\pi \ell! \, w_{0\ell}^{2(\ell+1)}}\exp{\left[-\frac{2r^2}{w_{0\ell}^2}\right]}, 
\end{equation}
where $P_0$ is the laser power and $w_{0\ell}$ is the beam waist. 
We use the single-ringed LG$_{0\ell}$ beam, with a radial mode index $p=0$ and a 
dark spot at the optical axis, where $\ell$ is the azimuthal index~\cite{LGBeams_zhang2017}. 
In this configuration, the plasma trapped in the dark core of the LG$_{0\ell}$
beam lacks a restoring force along the beam axis. 
This force can be introduced by adding plugging beams to the LG$_{0\ell}$
setup~\cite{kuga1997novel}. 
Alternatively, intersecting multiple LG$_{0\ell}$ beams can create a 
near-spherical (bicylinder) trapping volume, a three-dimensional single-beam 
configuration can also be used as described in~\cite{chaloupka1997single, chaloupka1999observation}, or overlapping optical cavities can be employed~\cite{cai2020monolithic}. 
For simplicity, we use a spherically symmetric LG potential, so 
in~(\ref{e_LGbeam}), the distance is $r=\sqrt{x^2+y^2+z^2}$.

\begin{figure}[t]
	\includegraphics[width=\columnwidth]{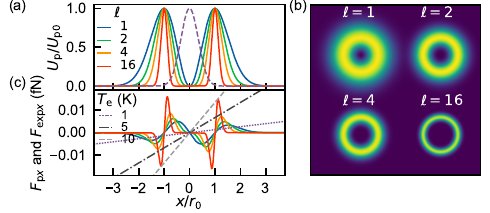}
	\caption{Optical ponderomotive potential shape and properties.
		(a)~Cross-sections of the ponderomotive potential for electrons normalized to 
		its peak value $U_{\text{p}0}$ for various azimuthal indices $\ell$. 
		The dashed curve shows the initial UNP density distribution with 
		$\sigma_0=30~\mu$m along one axis.
		(b)~Intensity profiles of the LG$_{0\ell}$ trap for selected $\ell$ values.
		(c)~Ponderomotive force on electrons and the force from electron thermal pressure. 
		Solid curves represent $F_{\text{p}x}$ for $\ell=\{16,4,2,1\}$ with fixed 
		potential depth $U_{\text{p}0}/k_\text{B}=20$~K; dashed curves show 
		$F_{\text{exp}x}$ for $T_{\text{e}}=\{1,5,10\}$~K.}
	\label{fig1}
\end{figure}

Figure~\ref{fig1}(a) shows the cross-sections of the ponderomotive potential 
$U_{\text{p}}$ normalized to its peak value $U_{\text{p}0}$. 
The maximum intensity of the LG$_{0\ell}$ beam occurs at $r_0 = w_{0\ell}(\ell/2)^{1/2}$. 
The peak potential $U_{\text{p}0}$ corresponds to the peak intensity $I(r_0)$. 
We vary $w_{0\ell} = \{124, 88, 62, 31\}~\mu$m for different $\ell = \{1, 2, 4, 16\}$
to keep the peak separation constant at $r_0 = 88~\mu$m. 
The peak intensities are many orders of magnitude below relativistic 
levels~\cite{RelativisticInt1}, and we do not consider the orbital angular 
momentum of the LG beams~\cite{denoeud2017interaction}. 
We assume that the particles are randomly distributed inside the trap volume 
with a Gaussian density distribution $n(\mathbf{r}) = n_0\exp{[-r^2/(2\sigma_0^2)]}$, 
where the initial size is $\sigma_0 = 30~\mu$m. The dashed line in 
Fig.~\ref{fig1}(a) represents the density profile of the UNP, normalized to the 
peak density $n_0 = N_{0\text{i,e}}/(2\sqrt{2}\pi^{3/2}\sigma^3)$.
Higher $\ell$ values enhance the force and enlarge the dark region at the trap 
center, making these configurations suitable for optical boxes due to their low 
scattering rate~\cite{NarutePhReviw2021}. Figure~\ref{fig1}(b) shows the 
intensity profile of the LG$_{0\ell}$ beam at the waist.

After initial thermal equilibration, UNP expansion is well described by 
hydrodynamic treatment~\cite{killian2007ultracold}. 
We estimate the per-ion force driving the ion expansion and compare it with the 
ponderomotive force acting primarily on electrons. 
The electron thermal pressure creates a per-ion force 
$\mathbf{F}{_\text{exp}}(\mathbf{r})=-k_{\text{B}}T_{\text{e}}\mathbf{\nabla} n(\mathbf{r}) / n(\mathbf{r})$~\cite{killian2007ultracold,gorman2021magnetic}. 
For simplicity, we neglect variations in electron temperature during UNP 
expansion~\cite{lyon2016ultracold,gorman2021magnetic}. 
Figure~\ref{fig1}(c) shows the ponderomotive force component acting on the 
electrons inside the trap, with $U_{\text{p}0}/k_{\text{B}} = 20$~K, compared 
to the electron thermal pressure force.
According to these simple estimates, the PT significantly affects plasma expansion when 
$k_{\text{B}}T_{\text{e}} / U_{\text{p}0} < 1$~\cite{smorenburg2013ponderomotive}.
However, plasma dynamics is more complex than this simple hydrodynamic model suggests.

The ideas presented so far have been tested numerically with the MD 
simulation method.
Standard kinetic theories failed to describe the strongly coupled plasma 
properties~\cite{killian2007ultracold}. 
One of the most effective ways to gain insight into UNP is through numerical 
modelling of non-equilibrium UNP using the MD method. 
We model the electron-electron and ion-ion interactions with a pure Coulomb potential. 
For electron-ion interactions, the Coulomb potential with repulsive 
core is used~\cite{RepCore}. 
Two-component plasma simulations were conducted in open-boundary conditions 
using open-source code LAMMPS~\cite{LAMMPS}.
The repulsion length scale is fixed at the value $\alpha=a_{\text{WS}}/80\approx60$~nm for all simulations. 
This length was chosen to be small enough to have no influence on the 
simulation results, as we verified by varying the repulsion length scale value between 30 and 100 nm.
The integration time step is $\delta t=0.2$~ps, which ensures that total energy is conserved to better than 1\%. 
We vary the initial number of ions and electrons 
$N_{\text{i}}=N_{\text{e}}=100...500$ in constant size $\sigma_0=30~\mu$m UNP 
to change the initial density in the PT.
The Debye screening length for all densities is $\lambda_{\text{D}}\ll\sigma_0$.
For the numerical simulations, we choose light lithium ions. 
Lithium is the closest element to hydrogen and is widely used for different 
laser cooling experiments with Rydberg atoms~\cite{Rydberg_aggregates}, 
photoionization phenomena study~\cite{LithFS2011,saakyan2022photoionization}, 
ion beam microscopy~\cite{IonMicroRev}, and quantum degenerate 
gases~\cite{NarutePhReviw2021}. 

\begin{figure}[t]
	\includegraphics[width=\columnwidth]{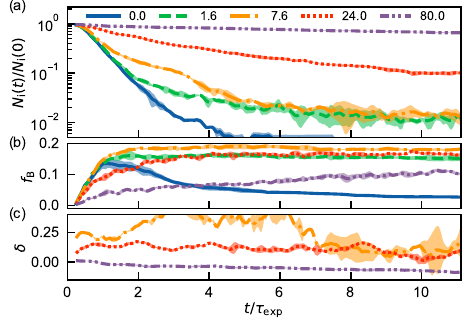}
	\caption{Time evolution of the normalized number of ions~(a), the fraction of 
	bound states~(b) and the charge imbalance~(c) inside the PT ($r<r_0$) for 
	different trap depths $U_{\text{p}0}/k_{\text{B}}$ (indicated in the legend in~K). 
	The initial number of particles $N_{\text{i,e}}(0)=500$, corresponds to an 
	initial peak density $n_0=1.18 \times 10^9$~cm$^{-3}$, $\ell=16$ and $T_e(0)=1$~K.
	Curves for $U_{\text{p}0}/k_{\text{B}}=0$ and 1.6~K in panel~(c) are 
	excluded for clarity due to high uncertainties.}
	\label{fig2}
\end{figure}

\begin{figure}
	\includegraphics[width=\columnwidth]{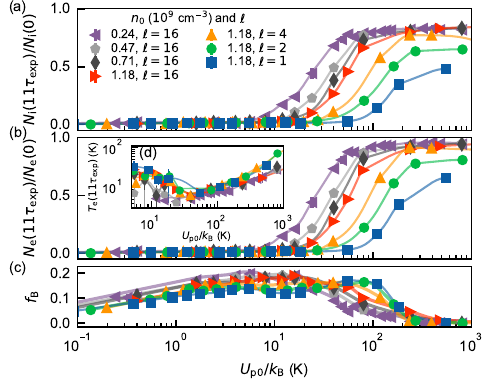}
	\caption{The normalized numbers of ions~(a), electrons~(b), and bound 
	states~(c) inside the trap volume, as well as the electron 
	temperature~(d), vs the trap depth $U_{\text{p}0}/k_{\text{B}}$
	after $t=11\tau_{\mathrm{exp}}$, for different initial UNP densities 
	$n_0$ and various trap shapes $\ell$.}
	\label{fig3}
\end{figure}

The number of ions inside the PT ($r<r_0$) vs time after photoionization is 
presented in Fig.~\ref{fig2}(a) for different values of $U_{\text{p}0}$ and 
$N_{\text{i}}=N_{\text{e}}=500$ corresponding to the initial peak density 
$n_0=1.18\times10^9$~cm$^{-3}$. 
According to the hydrodynamical description, the expansion time for light 
lithium ions is 
$\tau_{exp}=(m_{\text{i}}\sigma_0^2/[k_{\text{B}} T_{\text{e}}(0)])^{1/2}\approx0.9~\mu$s 
for initial plasma size $\sigma_0=30~\mu$m and $T_{\text{e}}=1$~K.
We investigate plasma properties inside the PT volume $r<r_0$. 
The expansion time without PT from this volume [the lowest curve in
		Fig.~\ref{fig2}(a)] is $\tau_{r_0}=0.83$~$\mu$s.

By analyzing the MD data, we identify the classical bound states according 
to~\cite{Recombination2011}. 
The electron is considered bound to the ion if the following criteria are met 
simultaneously: its energy drops significantly below the energy of the free 
electron; the rotation angle around the ion, integrated for a selected time 
interval $10^6\,\delta t$, is more than some critical angle.
After manually analyzing the trajectories, we introduced the final criterion that the electron-ion distance remain below $50\alpha$ over the same time interval and set the critical rotation angle to $8\pi$.

The time evolution of the fraction of bound states inside the trap volume is 
presented in Fig.~\ref{fig2}(b). 
The number of bound particles inside the trap ($r<r_0$) is normalized to the initial particle number $N_{\mathrm{i,e}}(0)$.
We identify bound particles and separate their 
contribution to all plasma parameters~\cite{RepCore}. Without PT, a sufficient 
number of electrons escape from the plasma after ionization. The charge 
imbalance described as $\delta=(N_{\text{i}}-N_{\text{e}})/N_{\text{i}}$
decreases with PT peak depth [see Fig~\ref{fig2}(c)], along with an increase in 
the number of trapped electrons and ions. A shaded region in the figure 
represents the calculation uncertainties extracted as the standard deviation of 
the few independent simulations with different initial random configurations and 
different random velocity distributions resulting in $T_{\text{e}}=1$~K and 
$T_{\text{i}}=1$~mK. The bound state fraction $f_{\text{B}}$ is the same for the 
$\alpha=30$~nm and 100~nm. There is no difference found 
within the uncertainties.

The trapping efficiency depends mainly on the UNP temperature.
The overall trapping efficiency depends on the temperature settled in the UNP 
after the equilibration of electrons and ions due to the 
DIH~\cite{murilloPRL2001,lyon2016ultracold,murphy2016disorder,killian2007ultracold}. 
This settled temperature strongly depends on the initial density of the UNP. 
In Fig.~\ref{fig3}, $N_{\text{i}}$, $N_{\text{e}}$, and $f_{\text{B}}$ at the 
time step $t=11\tau_{\mathrm{exp}}$ are shown for plasmas with different initial peak densities 
and different PT configurations $\ell$. 
Obviously, the trapping efficiency increases with the trap depth, especially for 
electrons (the charge imbalance $\delta$ in Fig.~\ref{fig2}(c) becomes negative, 
i.e. the number of electrons is higher than that of ions). 
When the UNP temperature becomes comparable to the trap depth, the plasma starts 
to confine inside the PT. 
In Fig.~\ref{fig3}(d), the kinetic temperature of electrons depending on the 
peak trap depth is shown. 
At higher $U_{\text{p}0}$, the formation of the bound-state fraction is 
suppressed [see Fig.~\ref{fig3}(c)] due to the increase in the electron kinetic 
temperature and the oscillation frequency of the electrons inside the potential 
wells. 
Even for low trap depth $U_{\text{p}0}/k_{\text{B}}\ll10$~K, the bound states 
formed in the UNP are successfully confined inside the PT volume.

For a single electron in a PT with $\ell=1$, if the oscillation amplitude is small, 
the ponderomotive potential is approximately parabolic, and the oscillation 
frequency is $\omega_\text{e}=[2P_0/(c m_e^2\pi w_{01}^4\varepsilon_0\omega^2)]^{1/2}$~\cite{moore1992confinement}. 
Numerical simulations for $\ell=1$ with small amplitudes are in excellent agreement 
with this calculated frequency $\omega_\text{e}$.

The coupling strength of the UNP is quantified by the coupling parameter 
$\Gamma_{\text{e,i}} = e^2/(4\pi\varepsilon_0 a_{\text{WS}}k_\text{B}T_{\text{e,i}})$, 
which represents the ratio of the Coulomb potential energy at the average 
interparticle spacing $a_{\text{WS}}$ to the average kinetic energy.
For a UNP confined in the PT, the ion coupling parameter lies in the range $\Gamma_{\mathrm{i}}\approx1\text{--}3$, while the electrons remain weakly coupled with $\Gamma_{\mathrm{e}}\approx0.1\text{--}0.3$ (values vary with trap depth). 
These coupling parameters are comparable to those reported in typical UNP experiments~\cite{killian2007ultracold}.

\begin{figure}
	\includegraphics[width=\columnwidth]{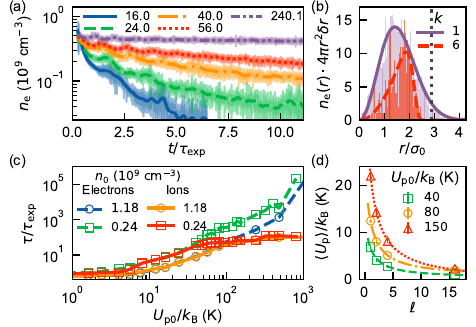}
	\caption{
	(a) Peak density $n_{\text{e}}(t)$ for several trap depths 
	$U_{\text{p}0}/k_{\text{B}}$ (in K, see legend). 
	(b) Radial density profiles at $t=0$ and $t\!\approx\!11\tau_{\mathrm{exp}}$
	for $U_{\text{p}0}/k_{\text{B}}=56$~K. 
	During equilibration the initially Gaussian profile ($k=1$) evolves toward a 
	flat-top shape ($k=6$) with damped oscillations; the dotted vertical line 
	corresponds to $r_0$ value. 
	(c) The lifetime $\tau$ of electrons (dashed line) and ions (solid line) inside 
	the PT versus $U_{\text{p}0}/k_\text{B}$ for two different initial densities. 
	For panels (a)--(c) $\ell=16$ azimuthal index is used. 
	(d) Density-weighted mean ponderomotive energy per electron 
	$\langle U_{\text{p}0}\rangle/k_{\text{B}}$ vs $\ell$, points are 
	simulation results and solid lines are least-squares fits to $A/(\ell+1)+c$, 
	consistent with the thin-shell model expectation 
	$\langle U_{\mathrm{p}}\rangle \propto U_{\mathrm{p}0}/(\ell+1)$.
	}
	\label{fig4}
\end{figure}

Figure~\ref{fig4}(a) shows the time evolution of the peak electron density $n_{\text{e}}$ for several trap depths $U_{\text{p}0}/k_{\text{B}}$, at fixed $\ell=16$.
The plasma confined in the trap follows the inner shape of the PT. 
For higher $\ell$, the plasma forms a homogeneous density profile with exponential 
decay, similar to neutral gases in optical boxes~\cite{NarutePhReviw2021} or UNP 
from magnetic traps~\cite{warrens2021expansion}. 
Initially Gaussian-distributed particles in the PT evolve into a flat-top distribution 
$n_{\text{e,i}}(\mathbf{r})=n_{0\text{e,i}}\exp{[-(1/2)([x^2+y^2+z^2]/\sigma^2)^k]}$
with $k>1$. 
At $t=11\tau_{\mathrm{exp}}$ for $U_{\text{p}0}/k_\text{B} = 56$~K, the radial distribution 
is well described by a flat-top function with $k = 6$ [see Fig.~\ref{fig4}(b)].
Following creation in the PT, the expanding plasma encounters the potential walls and rapidly equilibrates ($t<5\tau_{\mathrm{exp}}$), exhibiting damped oscillations in the peak density and size $\sigma$ [Fig.~\ref{fig4}(a)].
We also observe that lower $\ell$ values correspond to higher amplitude oscillations 
with well-defined frequencies.

Figure~\ref{fig4}(c) shows the plasma lifetime $\tau$ in the PT as a function of trap depth $U_{\text{p}0}/k_{\text{B}}$ for two initial densities. 
The lifetime is obtained by fitting the post-equilibration decay of the ion and electron numbers $N_{\text{i,e}}(t)$, as illustrated in Fig.~\ref{fig2}(a). 
Ion confinement by the PT is weak. 
Ions are held primarily by the electron space-charge potential within the trap.
Accordingly, increasing the trapped population extends the ion lifetime, which at sufficiently large populations may approach the electron lifetime. 
Three-body recombination (TBR) heats the plasma by creating bound states and limits the lifetime in the PT. This heating channel is included in our computational model.
The concept of trapping electrons and using the resulting space-charge to confine ions has been demonstrated in~\cite{gorman2021magnetic,bergeson2022ultracold}.
The complementary situation is also possible.
If the ions are confined, or their expansion is slowed, the ion space-charge provides a deeper
and longer lived electrostatic well that confines electrons.
This route is only available for non-hydrogen species with strong and accessible ionic transitions, where ion laser cooling can reduce the ion temperature and suppress expansion~\cite{LaserCoolingUCP2019,Krasnov2009}.

In the simulations the PT acts through the cycle-averaged ponderomotive force, and the electron quiver motion in the optical field is neglected. 
Conventional radio-frequency PTs for non-neutral particles do not confine a neutral plasma because collisional absorption, or inverse bremsstrahlung (IB), causes rapid electron heating. 
UNPs in external RF and microwave fields have been used to study collisional heating~\cite{LuitenHeating2012,LuitenMWHeating2022,RobertsRF2013}. 
The IB heating rate depends on the ratio of the electron–ion collision frequency $\nu_{\mathrm{e,i}}$ to the drive frequency~\cite{IBfreqexp2023}. 
In typical UNP conditions~\cite{killian2007ultracold,LaserCoolingUCP2019,gorman2021magnetic,PRLSteadyState2024} one finds $\nu_{\mathrm{e,i}}$ far below the GHz range. 
As a result, even intense laser fields produce negligible IB heating.
Energy acquired by an electron in one half-cycle is returned in the next unless a collision interrupts the motion, and the low collision probability makes the net per-cycle heating negligible.

In conventional transport theory, the power absorbed per electron is $P_{\mathrm{ei}}\approx2\langle U_\mathrm{p}\rangle\nu_{\mathrm{ei}}$~\cite{LuitenCavityPRA2019},
where the density-weighted mean per-electron ponderomotive energy, $\langle U_p\rangle=\int U_\mathrm{p}(r)n_\mathrm{e}(r)d^3r/\int n_\mathrm{e}(r)d^3r$, is determined numerically from the MD data and is shown in Fig.~\ref{fig4}(d) for several $\ell$ values and for $U_{\mathrm{p}0}/k_{\mathrm{B}}=40, 80,$ and $150$~K. 
The trajectory-averaged potential matches the ensemble average, consistent with an ergodic electron subsystem in a stationary two-temperature plasma. 
In a flat-bottomed trap only electrons near the walls sample the optical intensity. This thin-shell sampling yields the scaling $\langle U_\mathrm{p}\rangle \propto U_{\mathrm{p}0}/(\ell+1)$, consistent with our MD results. 
For $\langle U_\mathrm{p}\rangle=2.5$~K this gives $P_{\mathrm{ei}}/k_{\mathrm{B}}\approx200$~K/$\mu$s. 
More bremsstrahlung-specific models for the optical regime predict strong high-frequency suppression of absorption, so the transport estimate overstates heating in our conditions~\cite{IBfreqexp2023,IBfreqcalc2024}.

We also perform full-field simulations of the PT with an oscillating electric field corresponding to $U_{\mathrm{p}0}=96$~K at density $0.24\times10^9$~cm$^{-3}$ after equilibration. 
No measurable heating is observed over $150$~ns, and the trapping efficiency matches the cycle-averaged calculations.
To map the frequency dependence, we simulate a uniform UNP with periodic boundaries in uniform external fields from $50$~GHz to $30$~THz, which includes the CO$_2$ laser frequency~\cite{IBfreqexp2023,IBfreqcalc2024}. 
For RF range external fields, the IB heating agrees with theoretical estimates. 
At optical frequencies near a CO$_2$ laser we find no measurable IB heating over the simulated timescales.
These calculations indicate that optical-frequency heating of low-density UNPs is negligible~\cite{DataRep}.

The novel PT holds significant potential for various applications.
It enables dual trapping of charged particles and Rydberg atoms in the dark region and can guide
UNP or antimatter plasmas, assisting collimation in modern ion and electron microscopy~\cite{LuitenPRL2023Microscope}. 
Ground-state particles can also be confined at intensity maxima by dipole forces,
but a far-off-resonant CO$_2$ optical dipole trap provides only a shallow
potential~\cite{DipoleTrapsReview2000}. 
For antihydrogen, photon scattering is negligible while recombination produces atoms hotter than the trap depth, so their subsequent capture is inefficient~\cite{antihydrogenSpectroscopy}.

Several technical challenges exist for the experimental realization of the PT. 
Modern CO$_2$ lasers typically provide cw output power around 20~kW, while 
conventional cold-atoms experiments use IR lasers with two orders of magnitude 
less power. 
One solution is to use an optical cavity to reach the required circulating power level~\cite{cavityLIGO,cavityPulsed,HighPowerLG2013,cai2020monolithic}. 
In~\cite{cavityPulsed} a bow tie enhancement cavity demonstrated 500~kW average power at
1~\ensuremath{\mu}m wavelength. 
This corresponds to a peak ponderomotive energy of about 4~K. 
This depth provides only a modest enhancement of plasma electron confinement, but bound 
states are efficiently confined even at low depths as shown in Figs.~\ref{fig2}(b) and
\ref{fig3}(c). 
This enables accumulation and guided extraction of recombined products from antimatter plasmas,
similar to traveling-lattice antihydrogen beams~\cite{madsen2021formation}. 
Efficient coupling into a high order LG cavity eigenmode can also be challenging because 
coupling efficiency and mode purity are sensitive to alignment and aberrations. 
To our knowledge, megawatt average power enhancement cavities operating directly on high 
order LG modes have not yet been demonstrated, so overlapping optical cavities are a 
promising direction~\cite{cai2020monolithic}. 
Tight focusing limits trap volume and the trapped electron number, which shortens ion lifetimes,
so elongated cylindrical geometries are preferable for larger and lower density samples. 
Hybrid schemes that combine PT with magnetic confinement could suppress electron losses 
along field lines with reduced optical power~\cite{gorman2021magnetic} and may enable access
to strongly coupled, liquid- or crystal-like UNP regimes~\cite{Dubin1999}.

Finally, we note that optical dipole forces can be used to confine ions directly in 
species that have strong and accessible ionic transitions~\cite{iontrap2010}.
For UNPs of Ca$^+$ or Sr$^+$, a near resonant ion dipole trap with detuning of order few GHz
can reach $U_0/k_\mathrm{B}\sim 1$~K with sub watt optical power for $w_0\sim 30~\mu$m.
At these power levels, additional ponderomotive forces from the trapping beam are negligible
compared with the ion dipole potential.
The achievable confinement time is then limited mainly by photon scattering and recoil heating.
Scattering can be reduced by using blue detuned box like geometries.

In this work, we have proposed and analyzed a hollow-beam ponderomotive trap for ultracold neutral plasmas and demonstrated, via simulations, confinement of a two-component lithium plasma and Rydberg atoms within a dark, flat-bottom region. 
The combination of thin-shell sampling and optical drive frequencies suppresses inverse bremsstrahlung absorption.
Full-field simulations confirm the absence of measurable heating at the CO$_2$ frequency and validate the cycle-averaged description. 
Ion lifetimes increase with trapped population and approach electron lifetimes at large populations. 
These results establish a practical parameter window for the flat-bottom optical confinement of UNPs and outline routes to improve performance using cavity enhancement, high-order LG modes, elongated geometries, and hybrid optical-magnetic trapping. 
This approach enables simultaneous trapping of free charges and their Rydberg products, potentially advancing antimatter plasma trapping and production.

The supporting data for this Letter including MD trajectories and necessary 
LAMMPS input are openly available in the data repository~\cite{DataRep}.

\begin{acknowledgments}
	The author wishes to thank A.A. Bobrov, B.V. Zelener, B.B. Zelener, 
	V.A. Sautenkov, E.V. Vikhrov and S. Ya. Bronin for very useful discussions. 
	This work was supported by the Russian Science Foundation, Grant No. 23-72-10031 
	and by the Ministry of Science and Higher Education of the Russian Federation 
	(State Assignment No. 075-00270-26-00) with respect to providing the 
	computational resources.
\end{acknowledgments}
\bibliography{PTrap.bib}

\end{document}